\documentclass{aa}
\usepackage{graphicx}
\begin{document}
\newcommand{\kms}{{\rm km~s}^{-1}}
\newcommand{\ho}{{\rm km~s}^{-1}~{\rm Mpc}^{-1}}
\title{An Optical Time-Delay for the Lensed BAL Quasar HE~2149$-$2745
\thanks{Based on observations made with the Danish 1.5-m telescope (ESO,
La Silla, Chile) and at VLT UT1 Antu (ESO-Paranal, Chile) (Proposals: 
64.O-0205(B), 65.O-0214(B), 66.A-0203(B), 67.A-0115(B), 66.B-0139(A) 
and HST archives ID 8175).}}

\authorrunning{Burud et al.}

\author{I. Burud \inst{1,2} 
\and F. Courbin \inst{1,3,4}
\and P. Magain \inst{1}
\and C. Lidman \inst{5}
\and D. Hutsem\'ekers \inst{5}\thanks{Also Research Associate 
FNRS at the University of Li{\`e}ge, Belgium}
\and J.-P. Kneib  \inst{6}
\and J. Hjorth \inst{7}
\and J. Brewer \inst{5}
\and E. Pompei \inst{5}
\and L. Germany \inst{5}
\and J. Pritchard \inst{5}
\and A. O. Jaunsen \inst{5}
\and G. Letawe \inst{1}
\and G. Meylan \inst{2}}

\offprints{burud@stsci.edu}

\institute{Institut d'Astrophysique et de G\' eophysique, Universit\' e de
  Li\`ege, Avenue de Cointe 5, B--4000 Li\`ege, Belgium
\and
Space Telescope Science Institute, 3700 San Martin Drive, 
Baltimore, MD 21218, USA
\and Universidad Cat\'olica de Chile, Departamento de 
              Astronomia y Astrofisica,
              Casilla 306, Santiago 22, Chile
\and
GEPI, Observatoire de Paris-Meudon, Place Jules Janssen, F-92195
Meudon, France
\and
European Southern Observatory, Casilla 19, Santiago, Chile
\and
Observatoire Midi-Pyr{\'e}n{\'e}es (LAS), 
UMR 5572, 14 avenue E. Belin, 31400 Toulouse, France
\and
Astronomical Observatory, University of Copenhagen, Juliane
  Maries Vej 30, DK--2100 Copenhagen \O, Denmark
}

   \date{}

\abstract{We present optical $V$ and $i$-band light curves of the
gravitationally lensed BAL quasar HE~2149$-$2745. The data,
obtained with the 1.5m Danish Telescope (ESO-La Silla) between October
1998 and December 2000, are the first from a long-term project aimed at
monitoring selected lensed quasars in the Southern Hemisphere. A time
delay of $103\pm12$ days is determined from the light curves. In
addition, VLT/FORS1 spectra of HE~2149$-$2745 are deconvolved in order
to obtain the spectrum of the faint lensing galaxy, free of any
contamination by the bright nearby two quasar images. By cross-correlating
the spectrum with galaxy-templates we obtain a tentative redshift
estimate of $z=0.495\pm0.01$.  Adopting this redshift, a $\Omega=0.3$,
$\Lambda=0.7$ cosmology, and a chosen analytical lens model, our
time-delay measurement yields a Hubble constant of H$_{0}=66 \pm 8\,
{\rm km}~{\rm s^{-1}}~{\rm Mpc^{-1}}$ (1$\sigma$ error) 
with an estimated systematic error 
of $\pm3\, {\rm km}~{\rm s^{-1}}~{\rm Mpc^{-1}}$. Using non-parametric models
yields H$_{0}=65 \pm 8\, {\rm km}~{\rm s^{-1}}~{\rm Mpc^{-1}}$ 
(1$\sigma$ error) and
confirms that the lens exhibits a very dense/concentrated mass
profile. Finally, we note, as in other cases, that the flux ratio
between the two quasar components is wavelength dependent. While the
flux ratio in the broad emission lines - equal to 3.7 - remains constant 
with wavelength, the continuum of the brighter component is
bluer. Although the data do not rule out extinction of one quasar
image relative to the other as a possible explanation, the effect could
also be produced by differential microlensing by stars in the lensing galaxy.
}

\maketitle

\section{Introduction}

The time-delay between the gravitationally lensed images of a
distant source is a measurable parameter. Observed as the time
difference between the arrival dates of a single (lensed) wavefront
emitted by a distant source, it is directly related to the Hubble
constant H$_{0}$ (Refsdal~\cite{Refsdal}).  Obtaining accurate time-delay 
measurements in multiply lensed quasars can therefore yield {\bf
(i)} a determination of H$_{0}$ provided the mass distribution in the
lens is known, or {\bf (ii)} constraints on the mass distribution in a
given lens, using H$_{0}$ as inferred from other methods.  During the last
20 years much effort
has been devoted to the observations of lensed quasars, and in particular 
to long-term monitoring of selected
systems. Some of these are Q0957+561 (Schild \cite{Schild}, 
Vanderriest et al. \cite{Vanderriest}),
PG1115+080 (Schechter et al. \cite{Schechter}), B1608+656 
(Fassnacht et al. \cite{Fassnacht}) and
 B1600+434 (Burud et al. \cite{Burud} and Koopmans et al.  \cite{Koopmans}).
In this context, we have been  conducting a photometric monitoring
program at the Danish 1.5-m telescope at La Silla observatory
(ESO, Chile) since October 1998, with the goal of measuring the time-delays
in several well studied lensed quasars.  We present here the first result from
this program: the time-delay measurement in the two-image quasar
\object{HE~2149$-$2745}.

The lensed nature of the BAL quasar \object{HE~2149$-$2745} at $z=2.03$
was established by Wisotzki et al. (\cite{Wisotzki}).  This system
proves to be an easy target for monitoring at a site with reasonable
seeing conditions (up to 2\arcsec): it is bright ($B=17.3$)
and the two quasar images have an angular separation of
1.7\arcsec. The monitoring program, the light curves and the time
delay are discussed in Sect.~\ref{sect:data}, ~\ref{sect:phot} and
~\ref{sect:timedelay} below.

The lensing galaxy has been detected in HST NICMOS and WFPC2 images
but its redshift remains unknown.  With the aim of measuring this
redshift, we have obtained a spectrum of \object{HE~2149$-$2745} with
FORS1 at UT1 (ESO-Paranal, Chile). The analysis of the spectroscopic
data is described in Sect.~\ref{sect:spectro}. This section also
includes a discussion on the spectral differences between the two
quasar components.  Mass models and estimates of the Hubble constant
are presented in Sect.~\ref{sect:mass}.  Finally,
Sect.~\ref{sect:discussion} summarises the main results.



\begin{figure}
\centering
\caption{Field of view 5\arcmin x5\arcmin\, in size around 
\object{HE~2149$-$2745}. The three reference stars (labelled S1, S2, S3) 
used for the 
photometry and the two PSF-stars (PSF1, PSF2) used in the spectral 
deconvolution
are indicated. North is up and East to the left.}
\label{field} 
\end{figure}

\section{Photometric Monitoring
at the 1.5m Danish Telescope}
\label{sect:data}

\subsection{Observations and Data Reduction}

Weekly observations of \object{HE~2149$-$2745} were carried out at the
Danish 1.5-m telescope at ESO-La~Silla from October 1998 to December
2000.  The target is visible from the beginning of June  to the end
of December, which results in gaps of about $\sim$ 5 months in the
light curves.  Apart from these gaps, very few points were missed because of
poor weather or technical problems. Observations were
obtained in the $V$ and {\it Gunn $i$} bands with DFOSC (Danish Faint
Object Spectrograph Camera) which has a pixel size of 0\farcs395.
Fig.~\ref{field} shows the central region of the field of view of DFOSC,
i.e., 5 arcmin$^2$ of the 13 arcmin$^2$ available.  The $V$-band was given priority
because of the detector's better sensitivity at these wavelengths, but
observations were also carried out in the $i$-band in order to monitor
possible colour changes.  The exposure time was set to 900 sec during
the first 6 months of observations. It was increased to 1800 sec for
the rest of the observations in order to improve the photometric
quality for the faint B  quasar image.  For each observed data point
the total exposure time was divided into three dithered exposures in
order to remove cosmic rays and bad pixels.  The seeing varied from
0\farcs7 to 2\farcs9, with a median of 1\farcs6.
All the imaging data were pre-processed (bias-subtracted and flat-field
corrected) using standard IRAF routines.

\begin{figure}[t]
\centering
\caption{{\it Left}: Stacked $V$-band image of a 9\arcsec x 9\arcsec region
centered on \object{HE~2149$-$2745}. The seeing is $\sim$2\arcsec\, and the total 
exposure time is $\sim$ 17 hours. {\it Right}: Deconvolved image 
(FWHM$=$0\farcs4) obtained from the simultaneous
deconvolution of 57 frames. North is up and East is to the left.}
\label{decima}   
\end{figure}

\begin{table}
\caption[]{Positions of three reference stars relative to the
position of the A component in \object{HE~2149$-$2745}.}
\begin{tabular}{ccc} 
\hline
Star & R.A. (arcsec)    & DEC (arcsec) \\ 
\hline
S1 &    -48.126  &   136.268    \\
S2 &  -57.126     & -72.647       \\
S3 &  -38.160   &  -4.145    \\
\hline
\end{tabular}
\label{stars}
\end{table}


\begin{figure*}[t]
\centering
\includegraphics[width=8.7cm]{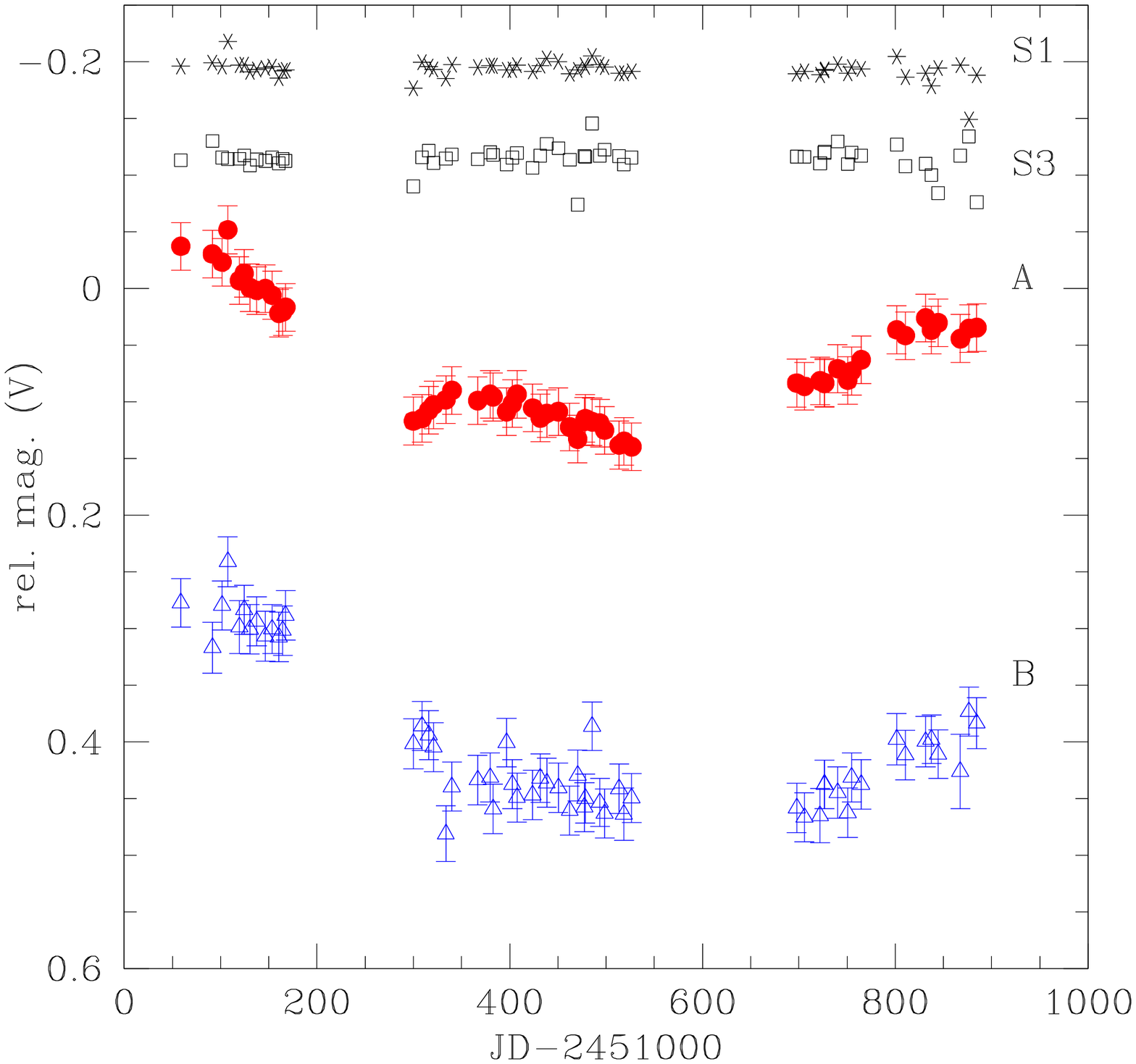}
\includegraphics[width=8.7cm]{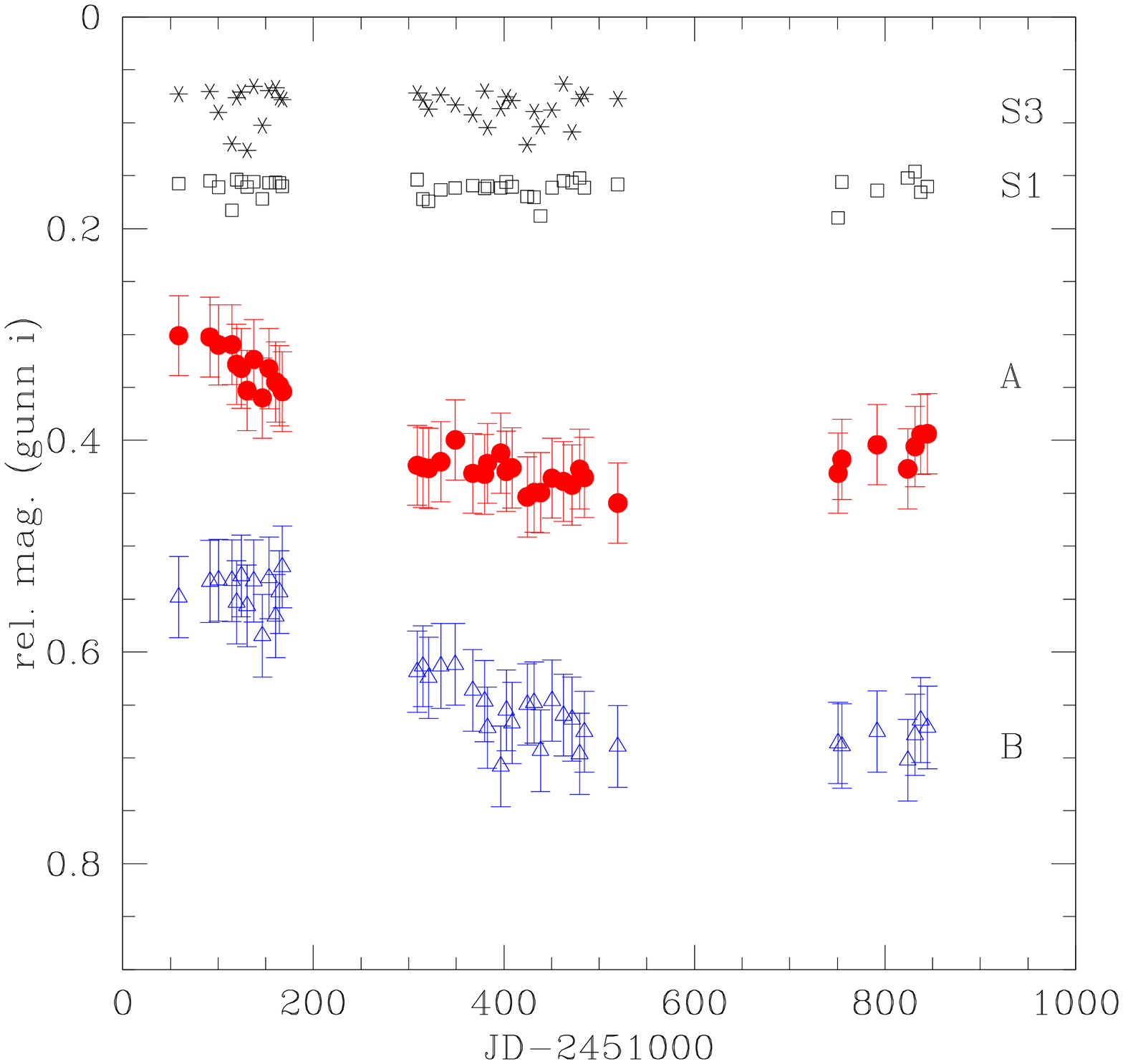}
\caption{Light curves of the two images A and B of \object{HE~2149$-$2745} and the comparison stars S1
and S3 (Table ~\ref{stars}).
The $V$-band light curves are displayed on the left while the {\it
Gunn i}-band light curves are shown on the right. The magnitudes of the quasar
and stars are calculated relative to the reference star S2.
The error bars include photon noise and PSF errors estimated from the
deconvolution of a reference star.  For display purposes, the B
component is shifted by $-1.3$ mag. and $-1.2$ mag. in the $V$ and {\it
Gunn i} band, respectively, with respect to their original values.
Likewise, S1 is shifted by $1.1$ and $1.35$, and S3 is shifted by 
$+1.15$ and $+1.35$ mag in the $V$ and {\it
Gunn i} filters, respectively. S3 was saturated in the last 
{\it Gunn i} images and is therefore not shown on the plot
for these frames.}
\label{lightcurves}
\end{figure*}


\begin{figure*}[t]
\centering
\includegraphics[width=12cm]{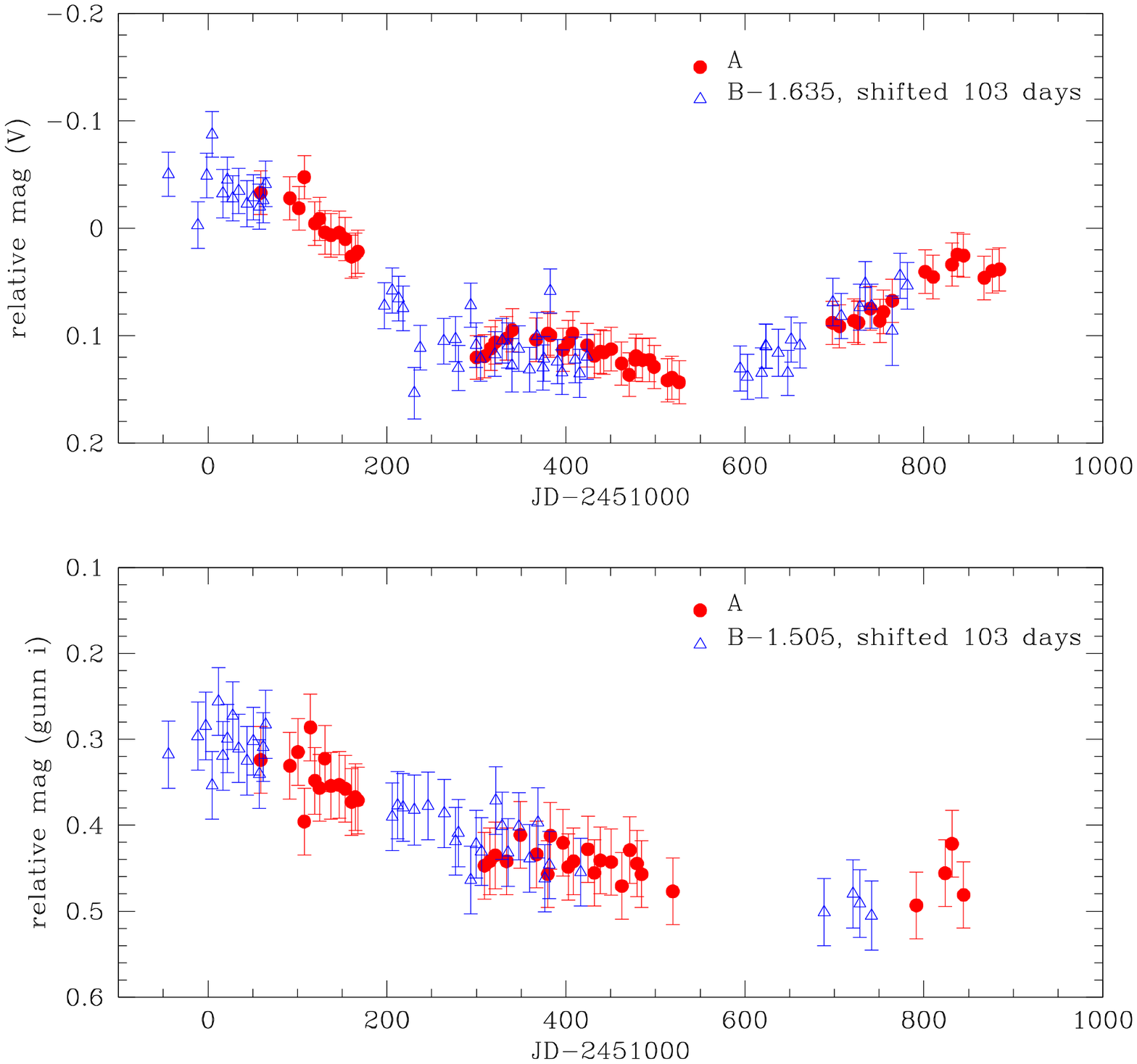}
\caption{The $V$ (top) and $i$-band (bottom) light curves where the B curve is
shifted by 103 days. The light curve of component B is shifted by 1.635
and 1.505 magnitudes in the $V$ and $i$-band, respectively, with
respect to their original values. The curves for the A component
are as in Fig. \ref{lightcurves}} 
\label{shift}
\end{figure*}


\section{Photometry}
\label{sect:phot}

The photometry of the blended quasar images was performed by applying
the MCS deconvolution algorithm (Magain, Courbin \&
Sohy~\cite{Magain}).  This algorithm has already been used to analyze
the data of several lensed quasars (e.g., Burud et al.~\cite{Burud},
Hjorth et al.~\cite{Hjorth}).  Its main advantage is its ability to
use all the data, even rather poor, irrespective of image quality
and lunar phase. The final deconvolved image is produced by
simultaneously deconvolving all the individual frames of the same object
from all epochs. The positions of the quasar images (two in the
present case) and the shape of the lensing galaxy are the same for all
the images and are therefore constrained using the total S/N of the
whole data-set.  The intensity of the point sources are allowed to
vary from image to image, hence producing the light curves.

The two  quasar components are  well separated in  our deconvolved
image  of  \object{HE~2149$-$2745}   (see  Fig.~\ref{decima}),  but  the
lensing galaxy remains too faint  to be detected. This is fortuitous
since contamination by an extended  object would complicate the analysis.

The light curves of \object{HE~2149$-$2745} consist of 57 data points in
the $V$-band and 41 points in the {\it Gunn $i$} band, as presented in
Fig.~\ref{lightcurves}.  The points between JD~2451800 and JD~2451900 were
obtained just  after the installation of  the new chip  on DFOSC.  The
new chip does  not have problems of charge diffusion,  as the old chip
had,  and  consequently, the effective seeing   is   now   considerably  improved.
Our original PSF star, for which the ADU counts were always in the linear 
regime on the old CCD,  became saturated on the new chip.
Another PSF star
further away  from the target had to  be used on a  few frames before
the  exposure time was  adjusted to  the new  chip.  These  points are
therefore more noisy than the previous points.

The data are plotted relative to a reference
star common to all images, star S2 in Table~\ref{stars}.  Two other
 stars were deconvolved as well, in order to check the
relative photometry and to check for systematic errors (see
Table~\ref{stars}  for the positions of the
stars).  The error-bars include both photon noise and additional
systematic errors, e.g., PSF errors. The latter is estimated by using
one of the reference stars, as explained in Burud et
al. (\cite{Burud}).

\section{Time-delay measurement}
\label{sect:timedelay}

By sliding the  light curves across one another, one  can make a rough
"by eye" estimate  of the time-delay of $\sim$~100  days, with A leading
B. Using the $\chi^2$ minimization method described
in Burud et al. (\cite{Burud}), a  more objective value of $\Delta t =
103  \pm  12$ days  is  found from  the  $V$-band  light curves.   The
$i$-band data are noisier and  have fewer data points but the measured
time-delay, $\Delta t = 104 \pm 31$ days, agrees well with the delay derived
from the  $V$-band.  A simultaneous minimization of the $V$ and $i$  band
curves gave a $\Delta t = 109 \pm 22$ days. 
The  errors quoted here  are obtained  from Monte
Carlo  simulations of  1000 sets  of light  curves, assuming  that the
photometric   errors   are   uncorrelated   and  follow   a   Gaussian
distribution.  The $\chi^2$ minimization method was also  performed 
on 1000 sets of curves where 5 randomly chosen
data points were  removed from each set.  The results
from these simulations yielded a time-delay of $\Delta t = 101 \pm 30$ days
and $\Delta  t = 107 \pm  30$ days in $V$ and  $i$ respectively, confirming
that the time-delay  measurement is robust.  The magnitude differences
between the A and the B components are found to be $1.635\pm0.001$ mag
in $V$ and $1.505\pm0.003$ mag in the $i$-band.  This corresponds to 
flux  ratios of  4.51 and  4.00  in $V$  and $i$ bands respectively
Although no erratic changes are detected in the
light curves within the  measurement errors, microlensing on long time
scales may still be present  as suggested by the VLT spectra presented
in  Sect.~\ref{sect:spectro}).  
Given that  the
$V$-band data contain more points  and are less noisy than the $i$-band
data, we adopt  the $V$-band estimate of the time-delay as the best one:
$\Delta t = 103 \pm 12$ days (1$\sigma$ error).
Fig.~\ref{lightcurves} summarizes  the
photometric data  and Fig.~\ref{shift}  displays the $V$  and $i$-band
light  curves with  component  B shifted  by  103 days.

\begin{center}
\begin{figure}[t]
\includegraphics[width=8.7cm]{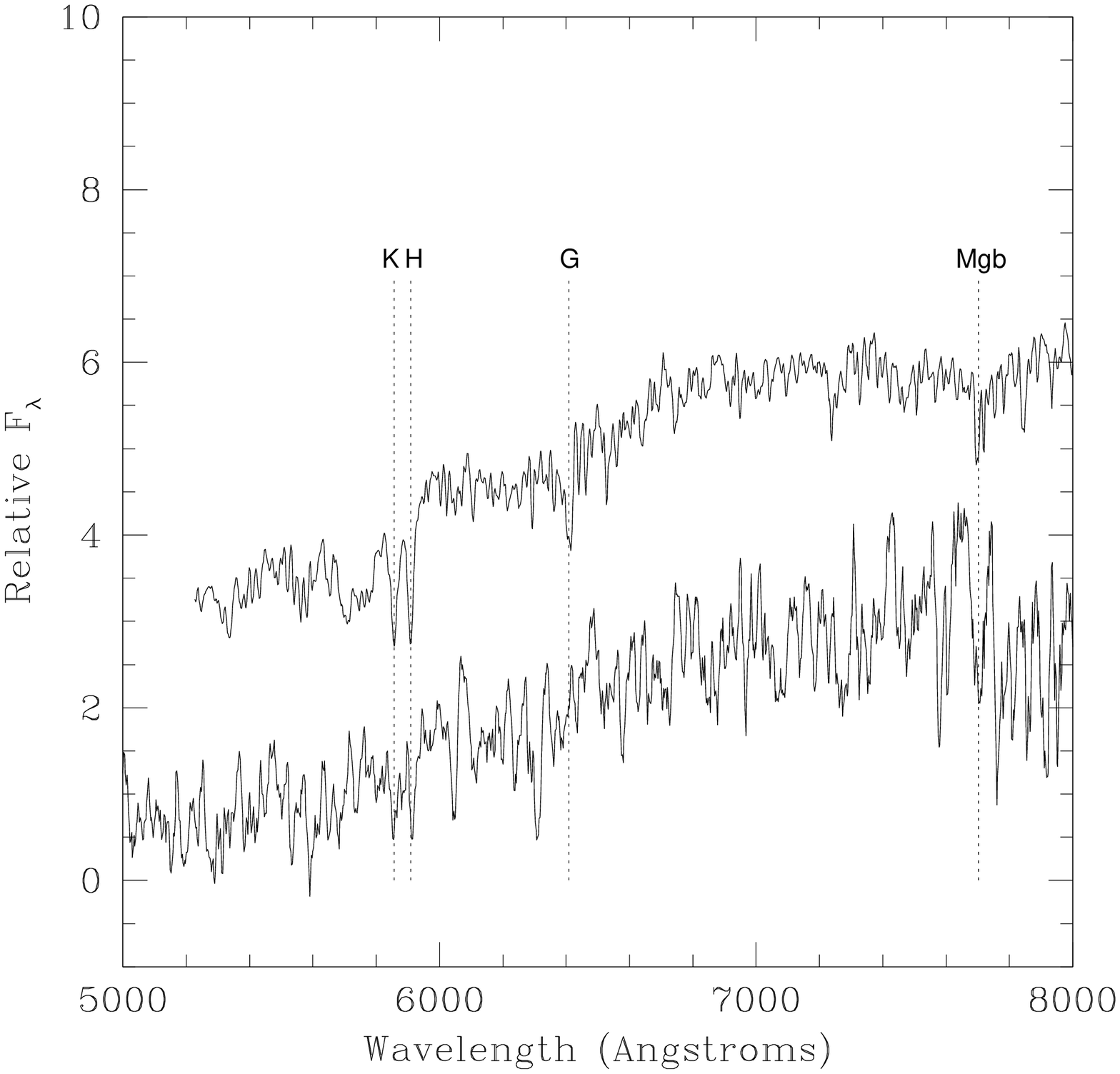}
\caption{Extracted spectrum of the lensing galaxy (bottom) and the template
spectrum from Kinney et al. (\cite{Kinney}) (top) used in the cross
correlation. The main spectral features, Calcium H and K lines, the
G-band and the Mg triplet, are indicated by  dashed
vertical lines. The lens redshift is tentatively $z=0.489$.}
\label{galaxy}
\end{figure}
\end{center}

\begin{center}
\begin{figure}[t]
\includegraphics[width=8.5cm]{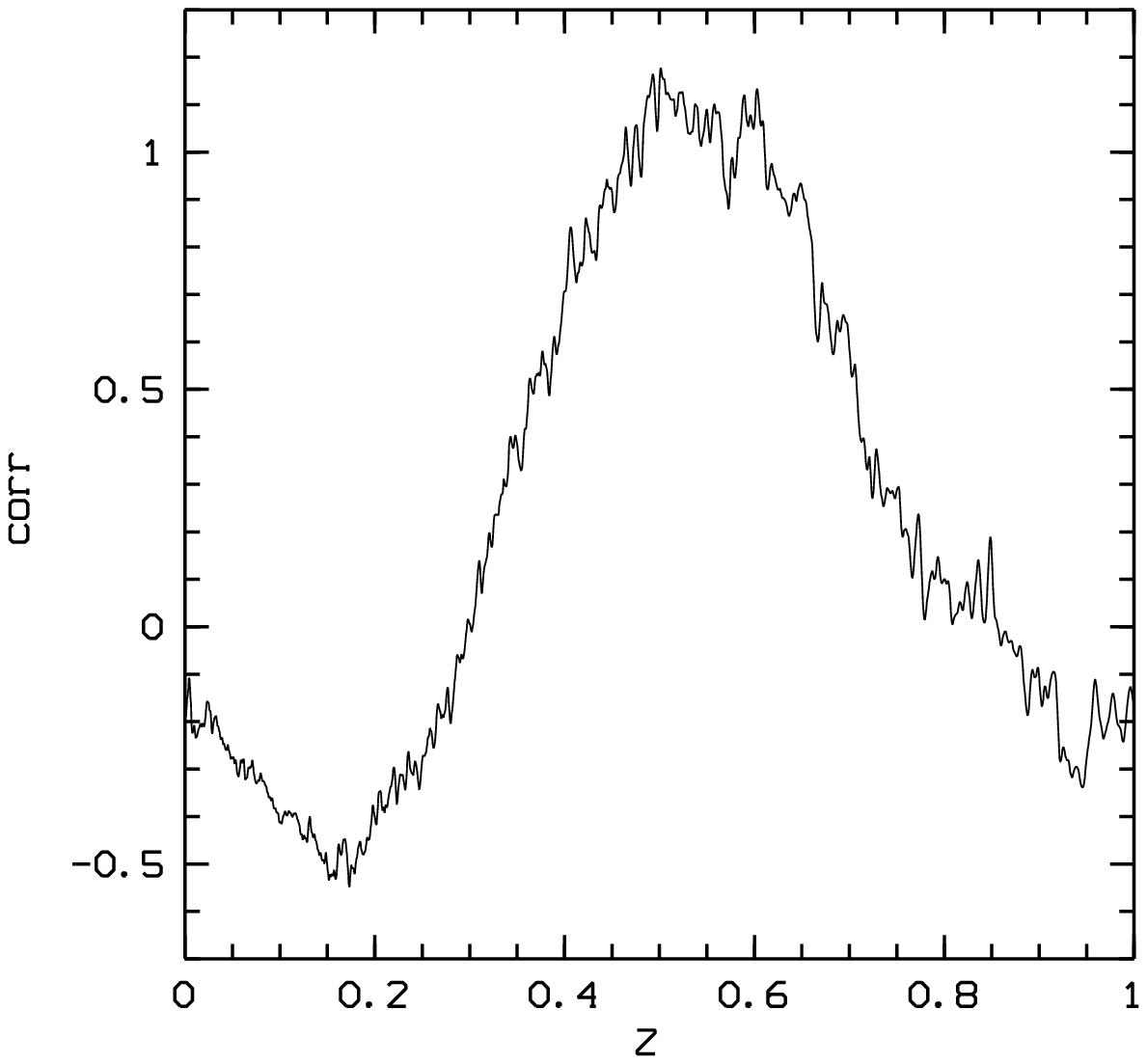}
\caption{Cross-correlation function of the galaxy spectrum
with the galaxy template from Kinney et al. (\cite{Kinney}). 
The centre of the correlation peak is at $z=0.535$. }
\label{corr}
\end{figure}
\end{center}

\begin{figure}[t]
\includegraphics[width=8.cm]{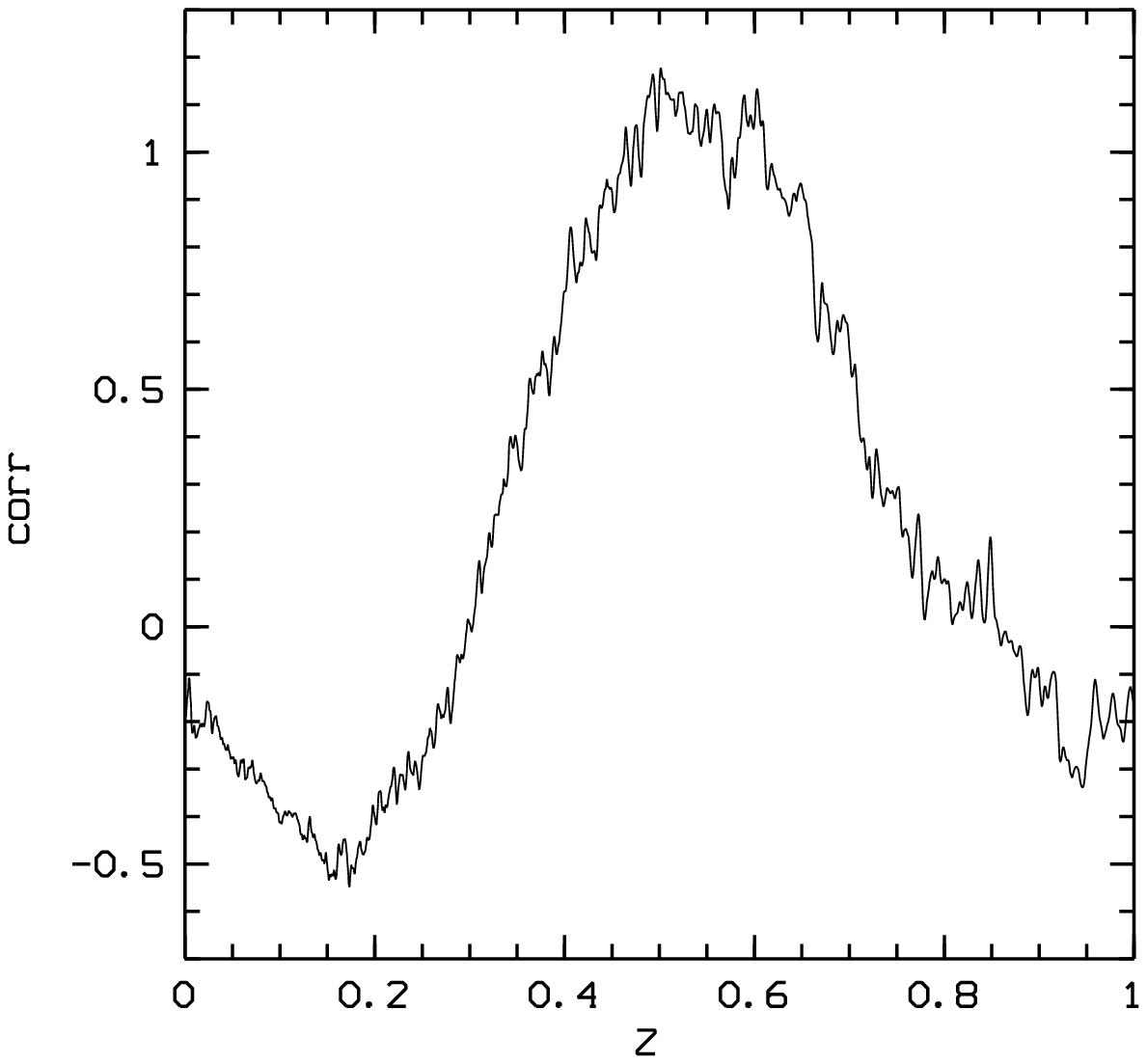}
  \caption{{\it Top}: spectra of the A and B components.  {\it
    Bottom}: the spectrum showing A$- 3.7 \times$B.  The scaling
    factor is chosen to best  cancel the emission lines. 
    Telluric absorption lines are indicated.}
  \label{specAB}
\end{figure}

\section{Spectroscopy of \object{HE~2149$-$2745}}
\label{sect:spectro}

\subsection{VLT spectroscopy}

To convert the time-delay into an estimate of the Hubble constant,
one must know the geometry of the lensing system and this includes the
redshift of the lensing galaxy. For this purpose, we took an
optical spectrum of \object{HE~2149$-$2745} with the ESO VLT/UT1 using
the Multi-Object-Spectroscopy (MOS) capability of FORS1.  The
observations were obtained on November 19, 2000 under fairly good
seeing conditions (0.8\arcsec). The 1\arcsec\, slitlets of FORS1 were
aligned to obtain simultaneously the spectrum of \object{HE~2149$-$2745}
and of 2 PSF stars (indicated in Fig.~\ref{field}) about as bright as
the two quasar images. The high resolution collimator was used,
resulting in a pixel size of 0.1\arcsec, in combination with the G300V
grism and GG435 order sorting filter.  Three exposures, each of 1000s,
were taken so that cosmic rays could be removed. 

\subsection{Redshift of the lensing galaxy}

Using the spatial information in the spectra of the two PSF stars, the
spectrum was spatially deconvolved with the spectral version of the
MCS deconvolution method (Courbin et al.~\cite{Courbin_a}).  The
deconvolution process decomposes the data into the individual spectra
of the two quasar images and the faint lensing galaxy (see
Figs.~\ref{galaxy} and ~\ref{specAB}), in a way similar to the one
described in Lidman
et al. (\cite{Lidman}) for HE~1104-1805.

\begin{center}
\begin{figure}
\resizebox{\hsize}{!}{\includegraphics{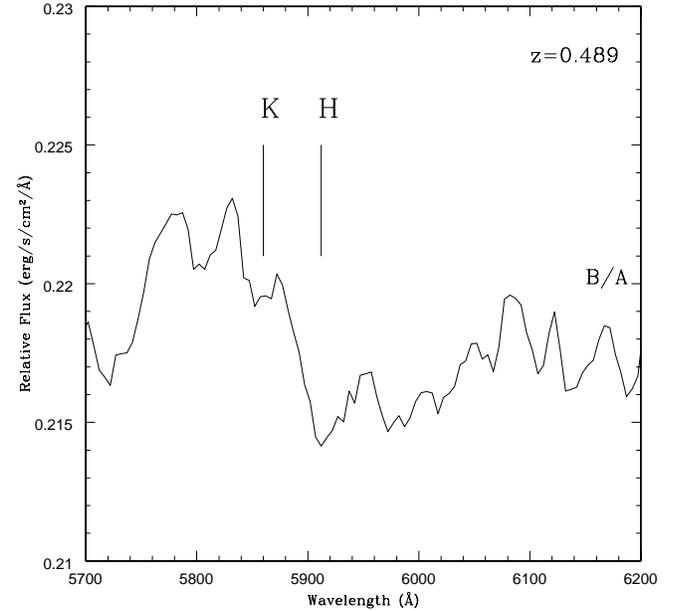}}
\caption{Spectrum of the B component divided by that of the A component
of \object{HE~2149$-$2745}. The plot shows a zoom on the region
corresponding to possible Calcium H and K absorption due to the lens at
$z=0.489$. The bump between 5750 and 5900 {\AA} is the residual CIII]+AlIII 
emission from the quasar.}
\label{bona}
\end{figure}
\end{center}
 
The lens spectrum was cross-correlated with a template spectrum of an
elliptical galaxy (Kinney et al.~\cite{Kinney}). 
Due  to the  low S/N the  calculated correlations  are low,
nevertheless, there is a significant ``bump'' in the correlation function
covering the redshift range $0.49\le z \le 0.60$ as shown in
Fig.~\ref{corr}. 
The spectrum of the lens and the
template are shown in Fig.~\ref{galaxy}, where the template has
been shifted vertically for clarity. The Calcium H and K lines, the
G-band and the Mg triplet are labeled.  
A measure of the reliability
of the redshift estimate is given by the r-statistic of Tonry and
Davis (\cite{Tonry79}), which is the signal-to-noise ratio of the main
peak in the cross correlation.  We find that $r=1.9$ meaning that the
signal-to-noise of the correlation peak is poor. A value above 3 is
considered secure (Kurtz \& Mink \cite{Kurtz98}). 

As can be seen on the HST image (cf. Fig.~\ref{HST}) the lensing galaxy 
lies very close to the B image. We therefore looked for
evidence of absorption in the spectrum of the background quasar.  In
order to remove the quasar spectral features, the B spectrum was
divided by that of the A component (Fig. \ref{bona}).  Two, possible
Ca H \& K candidates were found, one at $z=0.489$ and another at
$z=0.504$.  An integrated A+B spectrum at similar resolution but
covering shorter wavelengths has been published by Wisotzki et
al. \cite{Wisotzki}.  We looked for possible MgII absorption in the
quasar spectrum at the putative redshift of the lens.  Unfortunately,
the $z = 0.489$ line would fall in a strong absorption line of the
quasar, while a candidate at $z = 0.504$ is found in the wing of a
quasar emission.
The signal-to-noise ratio of the lens galaxy and the Ca H and K
absorption features in the B component are very low and we can not rule out any of the two estimates. We therefore take $z = 0.495\pm0.01$ 
to be the most likely redshift for the lensing galaxy.  
This is
within the estimated redshift range $0.37 \le z \le 0.50$ which is based
on the position of the lens in the fundamental plane (Kochanek et
al. \cite{Kochanek}).

\begin{center}
\begin{figure}[t]
\includegraphics[width=8.5cm]{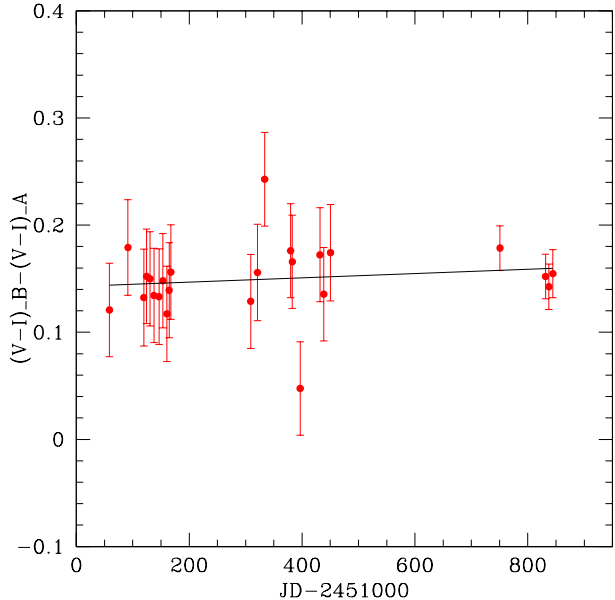}
\caption{Colour excess of the B component relative to the A component, 
$(V-I)_B - (V-I)_A$ as a function of time. No significant variations
can be detected over a period of 900 days. The solid line shows the
best fit line through the points.}
\label{fig:baplot}
\end{figure}
\end{center}

\subsection{Spectroscopy of the quasar images: 
colour differences, extinction or microlensing ?}

\begin{figure}[h]
\includegraphics[width=8.cm]{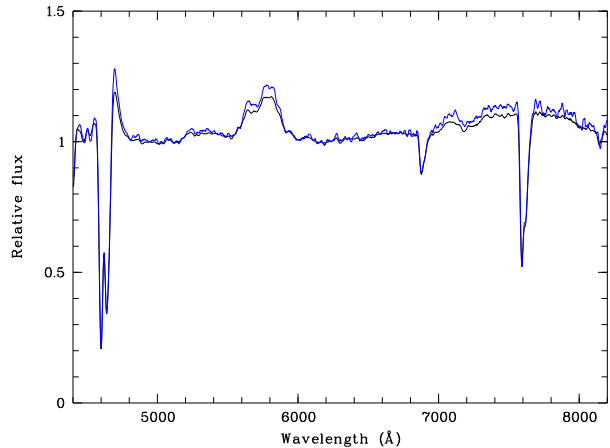}
  \caption{The A (thick) and B (thin) spectra of \object{HE~2149$-$2745} 
normalized to a common
continuum. The equivalent widths of the emission lines, including the FeII pseudo-continuum,
are smaller in A.  The equivalent width of the CIV broad
absorption line ($\sim$4650 \AA) is the same in both the A and B components.}
  \label{specnorm}
\end{figure}

The quasar spectra show that the continuum of component A is bluer than
the continuum of component B (Fig.~\ref{specAB}).  This confirms the
difference in flux ratio that was found from the light curves in the
$V$ and {\it Gunn i}-bands (Sect.~\ref{sect:timedelay}).  We note
three points of interest about these spectral differences:

\begin{enumerate}  

\item As shown in Fig.~\ref{fig:baplot} the colour difference between
the quasar images does not vary during the period of our observations
($\sim$ 3 years).

\item The emission line flux ratios at the date of the observation
are not wavelength dependent: all emission lines cancel in the
difference spectrum of Fig.~\ref{specAB}. This difference spectrum has
been produced by multiplying the spectrum of the B component by a {\it
constant} factor of 3.7 and by subtracting it from the spectrum of the
A component.  Fig.~\ref{specnorm} gives a different representation of
the same effect, where the spectra of the two quasar images are
normalized to the same continuum. The equivalent widths of all
emission lines (including the FeII pseudo-continuum) are smaller in A.

\item The CIV broad absorption line behaves as the continuum: its
equivalent width is unchanged in the A and B spectra
(Fig.~\ref{specnorm}).  

\end{enumerate}

The spectral differences cited above have essentially two
possible explanations: ({\it i}) differential magnification due to
microlensing, or ({\it ii}) differential reddening of the
B component by the lensing galaxy.

\subsubsection{Microlensing}

If microlensing is present in \object{HE~2149$-$2745}, one expects the
continuum region and the much more extended emission lines region
to be affected differently (Wambsganss \& Paczy{\`n}ski \cite{wambs}). 
In most AGN models, the size of
the continuum region depends on wavelength, and the Broad Emission Line
Region (BELR) is usually believed to be larger than the continuum
region by more than one order of magnitude.  Since the B component is
closest to the lensing galaxy we would expect that microlensing
effects occur most often in the B-component.  Microlensing of the A
component however is also  possible.

Assuming  the  lensed  quasar  in  \object{HE~2149$-$2745}  follows  the
``standard'' AGN model and that we are observing component A through a
network of caustics produced by  stars in the main lensing galaxy, one
can  imagine a  scenario where  the inner  -- and  bluer  -- continuum
region of  the component A is  being enhanced by a  larger amount than
the outer  redder parts. As the  BELR is much larger  than the central
AGN (even  the redder parts),  it remains unaffected  by microlensing.
This  interpretation  has already  been  proposed  to explain  similar
spectral differences  observed in the double  HE1104-1805 (Wisotzki et
al. 1993, 1995, Courbin et al. \cite{Courbin_b}).

In the case of \object{HE~2149$-$2745}
the Einstein 
radius of a typical deflector in the lens plane projected onto the
source plane is ($\Omega=0.3$,
$\Lambda=0$):

\begin{eqnarray}
R_{E}=\sqrt{\frac{4GM}{c^2}\frac{D_{ds}D_{s}}{D_{d}}}
=1.3 \times 10^{-3}(M/0.1M_{\odot})^{1/2} {\rm pc}
\end{eqnarray}
where $D_{d}$, $D_{s}$ and  $D_{ds}$ are angular diameter distances to
the   deflector,  the   source  and   between  deflector   and  source
respectively,  and $M$  is  the  mean mass  of  all microlenses.   The
duration  of  crossing the Einstein radius will be  up to  10 years  assuming a
relative velocity between source and lens that is equal or larger than
the  velocity   in  the   lensing  galaxy  which   is  of the   order  of
$\sim(100-400)  {\rm km  s^{-1}}$  (Mould et  al. \cite{Mould}). 
Although shorter microlensing events, such as a caustic crossings, 
can occur, the long time scale of crossing the Einstein radius shows
that an event may be present in our data, 
and  stable over our  relatively   short  period  of   observation  of  900   days,  as
Fig.~\ref{fig:baplot} would suggest.

\begin{figure}
\includegraphics[width=8.cm]{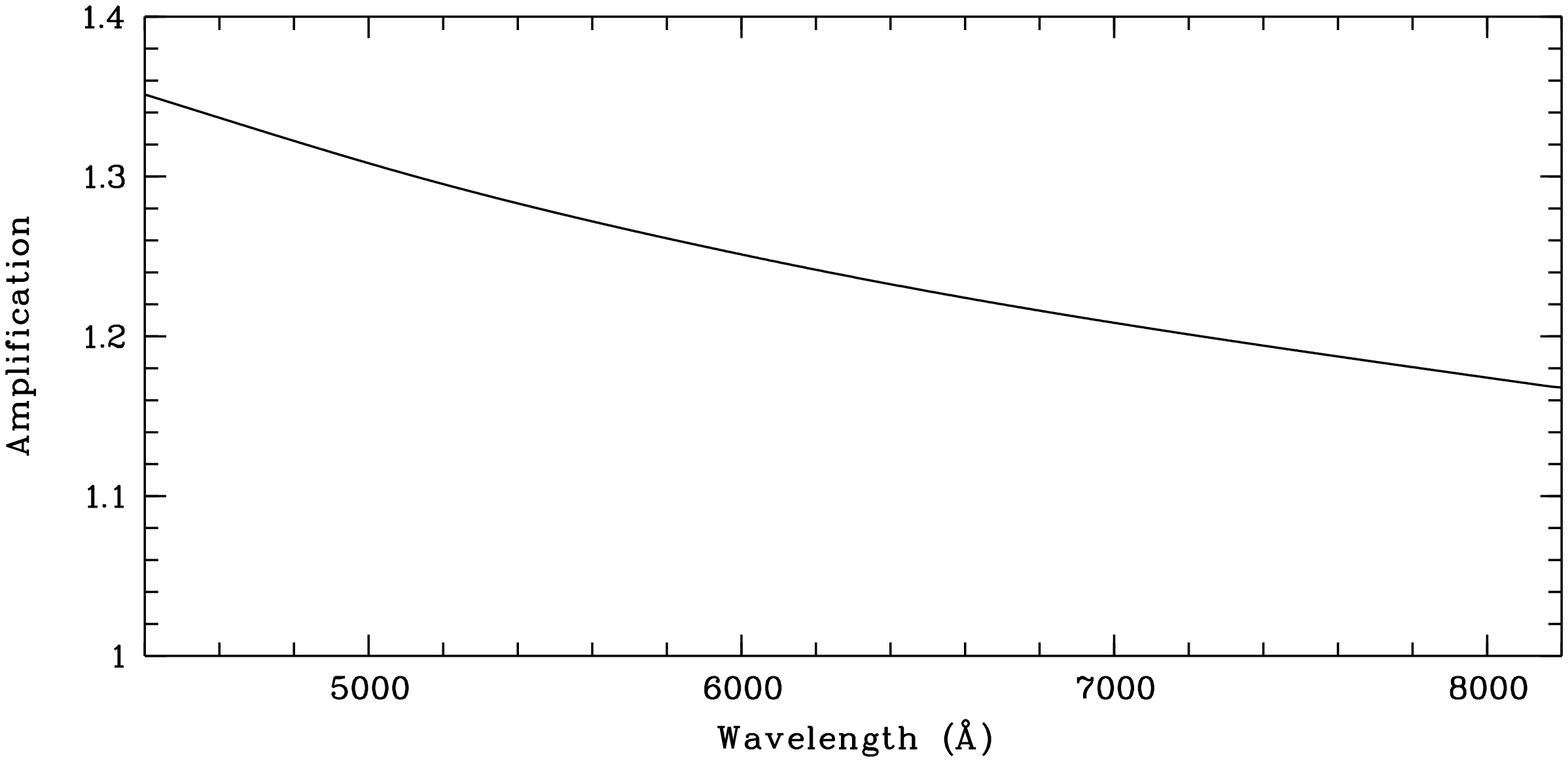}
  \caption{The residual magnification $\mu$ of A relative to B as a function of
  wavelength. This magnification factor is derived by fitting the A/B
  continuum ratio after correcting the B spectrum by the constant
  factor 3.7 (as in the lower panel of Fig.~\ref{specAB})}
  \label{Amlfig}
\end{figure}

\begin{figure}
\includegraphics[width=8.cm]{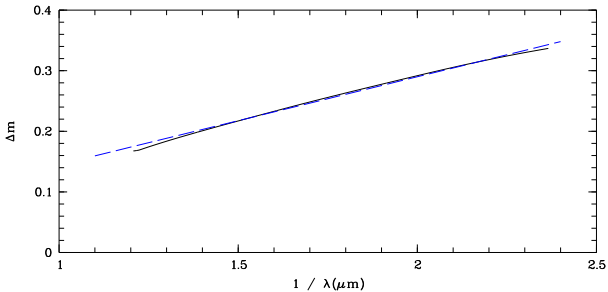}
  \caption{The magnification of component A in magnitude ($\Delta$m =
   2.5$\,\log \mu $) as a function of inverse wavelength (solid line). The wavelength
   dependence is well reproduced by $\Delta$m = 0.145$\, /
  \lambda(\mu$m )
  (dashed line)}
  \label{Amlfigl}
\end{figure}

Assuming that the continuum of A is amplified by microlensing, we may
derive the wavelength dependence of the magnification from the A/B
continuum ratio. This relative magnification factor $\mu$ is given in
Fig.~\ref{Amlfig}. It is derived by fitting the A/B continuum ratio
after correcting the spectrum of B by the constant factor 3.7. 
The relative magnification is directly related to
the size of the continuum region as a function of wavelength
(e.g. Schneider et al. 1992), the precise measurement of which
requires the knowledge of the background magnification $\mu_0$ due to
other distant subimages.  While $\mu$ can be estimated from
Fig.~\ref{Amlfig}, $\mu_0$ can only be obtained by measuring the background 
magnification before and after a caustic crossing event.
This can be done with a spectrophotometric monitoring. In
Fig.~\ref{Amlfigl}, we have plotted the magnification factor --
expressed as a magnitude difference -- as a function of the inverse
wavelength. It shows a remarkably tight wavelength dependence
strikingly similar to that found by Nadeau et al. (1999) for
Q2237+0305. This suggests that
\object{HE~2149$-$2745} presently suffers a chromatic microlensing event similar
(although of much longer duration) to the 1991 high-magnification event in
Q2237+0305 and which was interpreted -- on the basis of its colour
dependence -- as evidence for the thermal accretion disk model as the
source of UV-visible continuum emission in quasars (cf. Nadeau et
al. 1999).

\begin{figure}[t]
\includegraphics[width=8.cm]{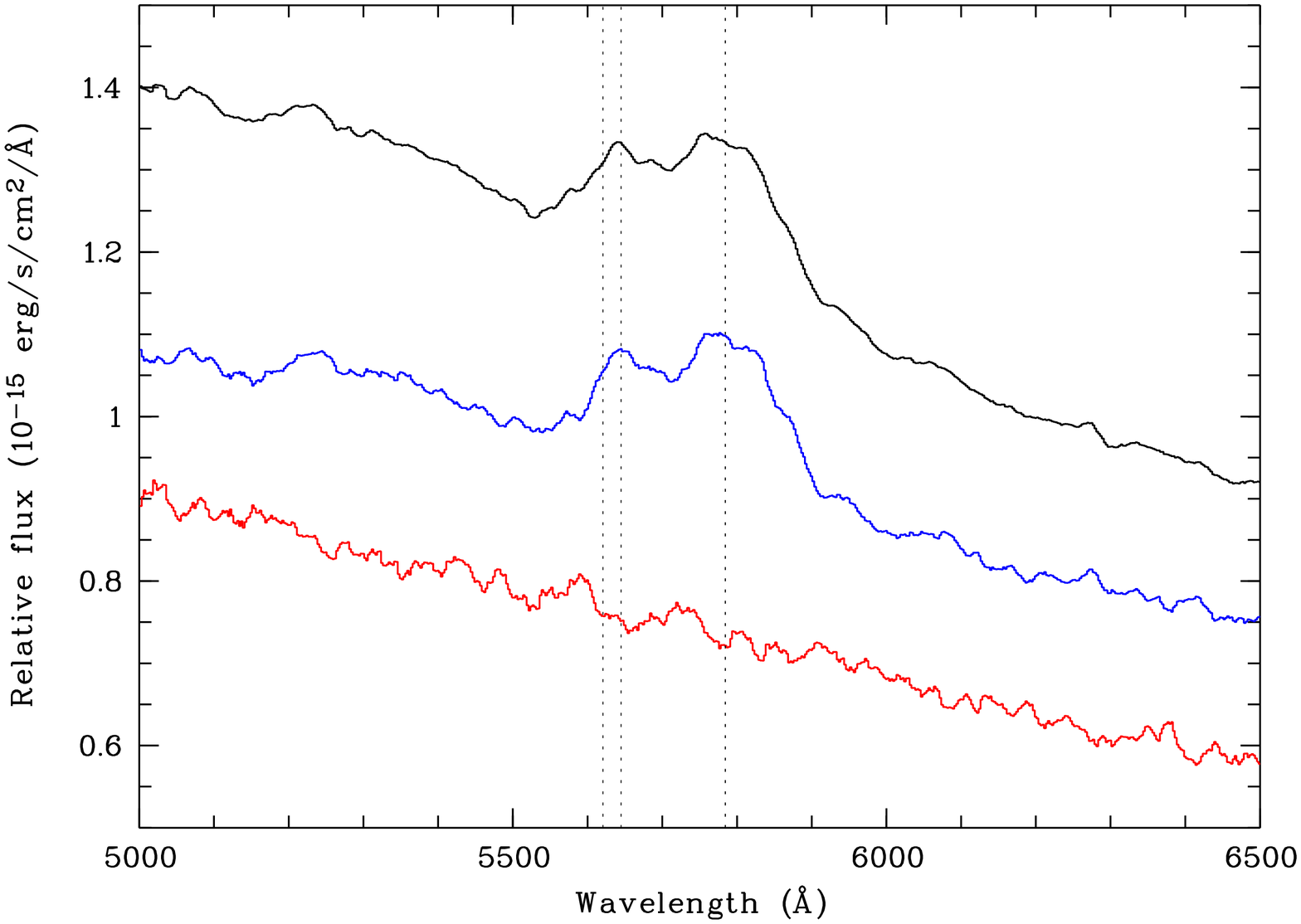}
  \caption{An enlargement of the spectra of components A (upper) and B
   (middle, multiplied by 3.7), together with the difference spectrum
   A$- 3.7 \times$B (lower, multiplied by 2). The region of the CIII]
   emission line is illustrated.  The wavelengths of AlIII
   $\lambda\lambda$1855, 1863 and CIII] $\lambda$1909 are indicated at
   the redshift of the quasar ($z$= 2.03).  This blend is typical of BAL
   quasars (Hartig \& Baldwin 1986). Some subtle differences in the blue
   part of the line profiles may be noticed (see text).}
  \label{Sdiffig}
\end{figure}

While broad emission lines should be essentially unaffected by the
caustic crossing, some subtle differences may arise in the line
profiles if some parts of the BELR are selectively magnified (Nemiroff
1988, Schneider \& Wambsganss 1990). Looking in detail at the line
profiles of components A and B (Fig.~\ref{Sdiffig}), small differences
are indeed present in the blue wing of the AlIII and CIII] emission
lines. Interpreted in terms of microlensing, this could indicate that
 part of the BELR is being magnified.  Similar profile variations
are predicted by the models of Schneider \& Wambsganss (1990).
Interestingly, no difference in the CIV BAL profile, like those reported for the 
quadruple BAL quasar H1413+117;
cf. Angonin et al. \cite{Angonin} and Hutsem\'ekers \cite{Hutsemekers}, 
is noticed.

The present findings may be of importance for investigating the inner
structure of quasars. Spectroscopic data obtained at regular time
intervals, e.g., spectra separated by intervals corresponding to the
time-delay, will test the microlensing interpretation, cancel possible
time-delay effects (see Wisotzki et al. 1995) and help to probe the
innermost region of quasar structure, using microlensing
magnification.

\subsubsection{Reddening}

An alternative explanation for the observed colour difference between
quasar images is that the B image is reddened by the lensing
galaxy. Assuming this is true, we can estimate the amount of reddening
which would correspond to the strength of the putative Ca H \& K
absorption lines seen in the B spectrum (Fig.~\ref{bona}), if average
Galactic conditions apply to the present case.  From the interstellar
line observations of Sembach et al.\ (1993), we derive a mean relation
between E(B--V) and the Ca K equivalent width.  With a de-redshifted
equivalent width $\sim 100$ m\AA, this would imply a colour excess
E(B--V) $\sim 0.05$, which corresponds to E(V--I) $\sim 0.1$, using
the extinction curves of Mathis (1990), after proper redshift
corrections are applied.  This is compatible with the measured colour
difference of 0.15 (Fig.~\ref{fig:baplot}).  The extinction can also be estimated by fitting the magnitude
difference of the A and B components as a function of
wavelength (Fig.~\ref{Amlfigl}) with a
typical extinction law. In the wavelength range of interest, the
Galactic and SMC extinction curves are very similar (Pei \cite{Pei})
and can be reasonably well represented by

\begin{equation}
A_{\lambda}=\frac{1.75E(B-V)}{\lambda(\mu m)}
\end{equation}
where $A_{\lambda}$ is the extinction in magnitudes, and where the
ratio of the total-to-selective extinction $R_{V}$ is taken to be
$R_{V}=3.1$.  Identifying the wavelength dependence of the magnitude 
difference  in
Fig.~\ref{Amlfigl} to this extinction curve at the redshift of the
lens, we have 1.75 E(B--V)(1+$z_{lens})=0.145$ i.e., E(B--V)$=0.05$
which is in  agreement with what is found from the Ca H \& K
lines.

In the case of extinction by dust in the lens we would expect the
continuum and the BELR to be equally affected. This does not seem to
be the case; however, the wavelength range between CIV and the
CIII]-AlIII complex is small. A more sensitive test of the dust
hypothesis will be to observe H-alpha in the infra-red, as this will provide
a larger wavelength baseline to measure differences.

On the basis of the available data, no interpretation of the apparent reddening
can 
be excluded,
although microlensing seems a somewhat more natural explanation for
the observed spectral differences.

\begin{figure}[t]
\includegraphics[width=8.5cm]{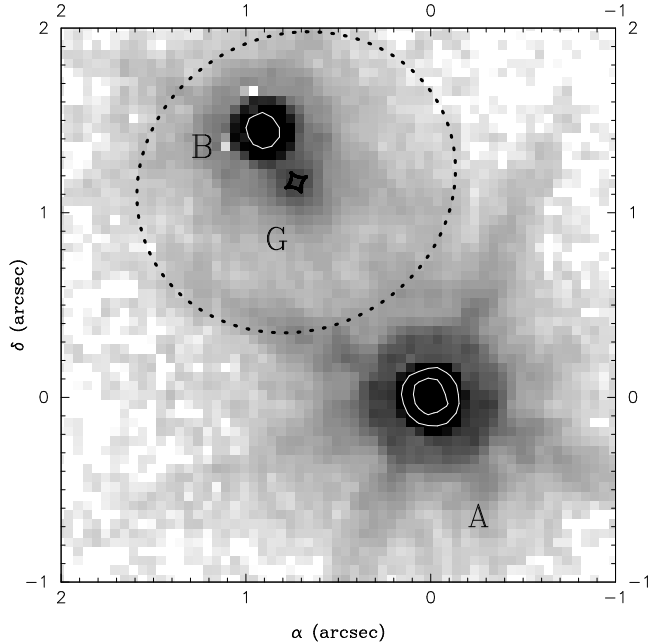}
\caption{Field of 3\arcsec\, around \object{HE~2149$-$2745}, where the
lensing galaxy is indicated as G, close to the faint quasar image
B. Also indicated are the critical and caustic curves created by a single
lensing galaxy and no external shear. The image has been obtained from
the CASTLEs public survey of gravitational lenses. This I-band image comes
from HST/WFPC2  (Kochanek et al.).}
\label{HST}
\end{figure}

\begin{figure}[h]
\includegraphics[width=8.5cm, angle=270]{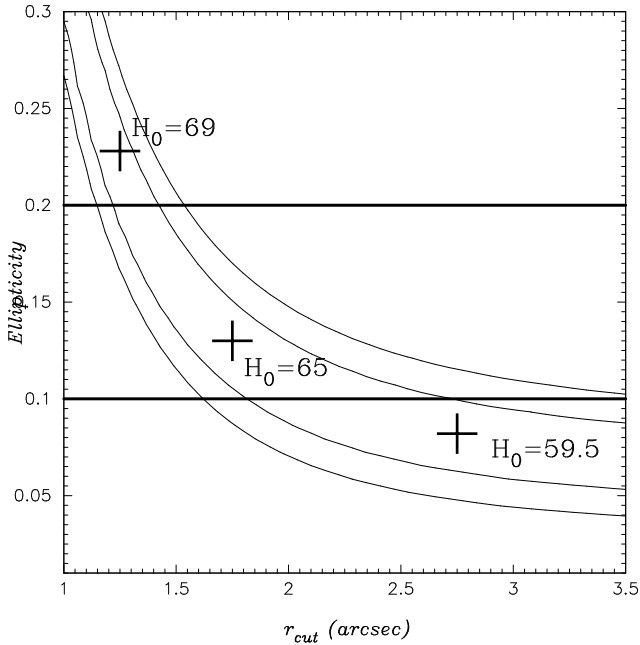}
\caption{Results of the model fitting when exploring the
ellipticity/cut radius space. The lines show the best models, i.e.,
they show the area where the models fit the data at 1$\sigma$ (inner
lines) and 3$\sigma$ (outer lines). Values for H$_{0}$ are given for
different models among the most likely. The horizontal lines give the
plausible range of ellipticities, given the HST images. The plot
displayed here uses $\Omega=0.3$, $\lambda=0.7$.}
\label{cut_ell}
\end{figure}

\section{Lens  modeling}
\label{sect:mass}

The time-delay measured for \object{HE~2149$-$2745} can be used to infer
an estimate of H$_0$, based on modeling of the total gravitational
potential responsible for the lensing effect. This includes the main
lensing galaxy and any intervening massive cluster along the line of
sight to the quasar. \object{HE~2149$-$2745} is a relatively easy case
to model, as there is no obvious mass concentration along the
line of sight to the source apart from the main lensing galaxy. Only a
marginally detected galaxy concentration is seen 30\arcsec\, West and
1\arcmin\, North of the main lens. In addition, the lensing galaxy is
almost aligned with the two quasar images (see Figs.~\ref{field} and
\ref{HST}). Two different approaches can be used to convert time
delays into H$_0$: models that involve an analytical form for the
lensing galaxy, and models that involve more degrees of freedom, e.g.,
a pixelated surface mass density (Saha \& Williams \cite{Saha}, Williams \& Saha,
\cite{will_saha}). We use both approaches.

\subsection{Analytical models}

Our analytic model uses the mass profile of Kneib et
al. (\cite{kneib96}): a truncated
Pseudo-Isothermal-Elliptical-Mass-Distribution (PIEMD).  Truncated
PIEMD are elliptical mass distributions smoothly truncated at radius
$r_{cut}$. For radii smaller (respectively larger) than $r_{cut}$, the
projected surface mass density profile is varying as $r^{-1}$
(respectively $r^{-3}$).  The interest in such profile is their
ability of characterizing any ellipticity in the mass distribution as
well as having a total finite mass.  We fit the model to the publicly
available 
HST/WFPC2 data (which offers better resolution and sampling than
NICMOS) obtained by Kochanek et al., and use the emission line flux 
ratio of 3.7
calculated from our VLT spectra. During the fit, the astrometry of the
quasar images relative to the lens is fixed, as well as the redshift
of the source and lens. The free parameters include the velocity
dispersion, ellipticity, which is here defined as 
$[a^2-b^2]/[a^2+b^2]$ where $a$
and $b$ are the long and short axes of the lens, and cut radius of the
lens.

A common  problem in  lens modeling is  the mass-sheet  degeneracy. In
other words, several mass profiles  and in particular mass profiles of
different compactness,  will reproduce  the same time-delay (Gorenstein et al. \cite{Gorenstein}).   One has
therefore to constrain the models using observational data and explore
the widest range of  physically acceptable models.  A first constraint
comes  from  the  HST  images:   the  lens  is  almost  round.   After
PSF-subtraction of the  quasar images we fit the  shape of the lensing
galaxy and obtain an ellipticity  of about 0.1--0.2. Assuming that the
mass  distribution  follows the  light  distribution  we can  restrict
ourselves to this range  of ellipticities.  Fig.~\ref{cut_ell} shows a
family of models exploring the  ellipticity vs. $r_{cut}$ space and we
see that little spread in the values of H$_0$ is observed in the range
of ellipticities  0.1--0.2.  We assume  ellipticity $\varepsilon=0.15$
which  implies $r_{cut}=1.6\arcsec$  and  H$_{0}= 66 \pm  8  ~\ho$ with  
additional systematic
errors of  $\pm~3~\ho$  due to the  limited range of  ellipticity.  We
also try to investigate the existence of a galaxy core.  For this we scan the
possible values for the core radius.  The lens model excludes any value
of $r_{core}$  larger than  0.02 arcsec. Within  this limit,  the core
radius has basically  no effect on the determination  of H$_0$, so the
choice of a given core radius  is not critical. In our fiducial model,
we use a core radius  of $\sim0.002$ arcsec which corresponds to $\sim$10 pc.
Using this, and keeping the other parameters free, we find a lens with
the   parameters  summarized   in  Table~\ref{model}.    We   show  in
Table~\ref{cosmo} the effect of a change in cosmology. Due to the 
relatively low redshift of the lensing galaxy the influence of a change
in cosmology is small compared to the measurement errors.

\begin{table}
\caption[]{Model parameters for \object{HE~2149$-$2745}. The values
between brackets are {\it fixed} during the fit, the others are
fitted. The core radius is negligible.}
\begin{tabular}{lc} 
\hline
Parameter     &  Fit  \\ 
\hline
$\Omega,\lambda$  & [0.3, 0.7] \\
Ellipticity       & 0.13 $\pm$ 0.05       \\
P.A. (degrees)           & 62 $\pm$ 5    \\
Vel. disp.  ($\rm {km\, s^{-1}}$)     & $198 \pm 10$     \\
$r_{cut}$         & [1.25\arcsec]         \\
Core radius       & [0.002\arcsec = 0.01 kpc]\\
time-delay        & [103 days]   \\
\hline
\end{tabular}
\label{model}
\end{table}
\begin{table}
\caption[]{Influence of the cosmology on the modeling
of \object{HE~2149$-$2745}. Values of H$_{0}$ are listed for 
an Einstein--deSitter Universe ($\Omega=1.0$, $\lambda=0.$), 
an open Universe  ($\Omega=0.3$, $\lambda=0.$) and a flat Universe
with non-zero cosmological constant  ($\Omega=0.3$, $\lambda=0.7$).
The model parameters are fixed
and set to the ones in Table~\ref{model}.}
\begin{tabular}{cr} 
\hline
Cosmology                 &  H$_0\, (\ho)$    \\ 
\hline
$\Omega=0.3,\, \lambda=0.7$  &  66 $\pm$ 8 \\
$\Omega=0.3,\, \lambda=0.0 $  &  67 $\pm$ 8 \\
$\Omega=1.0,\, \lambda=0.0 $  &  61.5 $\pm$ 8 \\
\hline
\end{tabular}
\label{cosmo}
\end{table}

If   we  take   the  currently most   popular  cosmology,   i.e.,  $\Omega=0.3$,
$\lambda=0.7$, and $\Delta t = 103 \pm 12$ days, we find H$_0 = 66 \pm
8\, \ho$ with an additional estimated systematic error of $\pm~3~\ho$ due to parameter choices.

\begin{figure}[t]
\includegraphics[width=8.5cm]{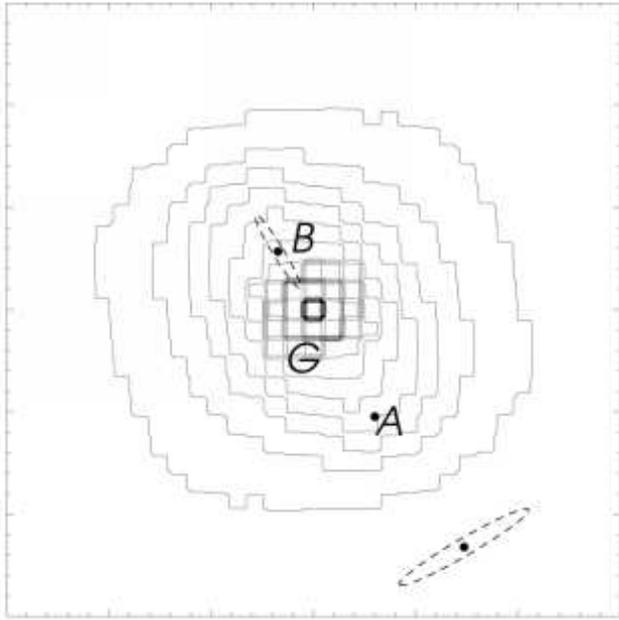}
\caption{Mass distribution found with the pixelated models with
cosmology $\Omega=0.3$, $\lambda=0.7$. Almost no
ellipticity is found. The contour levels represent  
$\kappa$ = 1/3, 2/3, 1, etc. 
The scale is 3\arcsec\, on a side and the
astrometry is the same as for the analytical models. The dashed
ellipses around the quasar images represent the magnification matrix,
i.e., the axes are proportional to the eigenvalues and the orientation
corresponds to the eigenvectors. The spot between the quasar images is
the position of the lensed source in the source plan.}
\label{mass_saha}
\end{figure}
\begin{figure}[t]
\includegraphics[width=8.5cm]{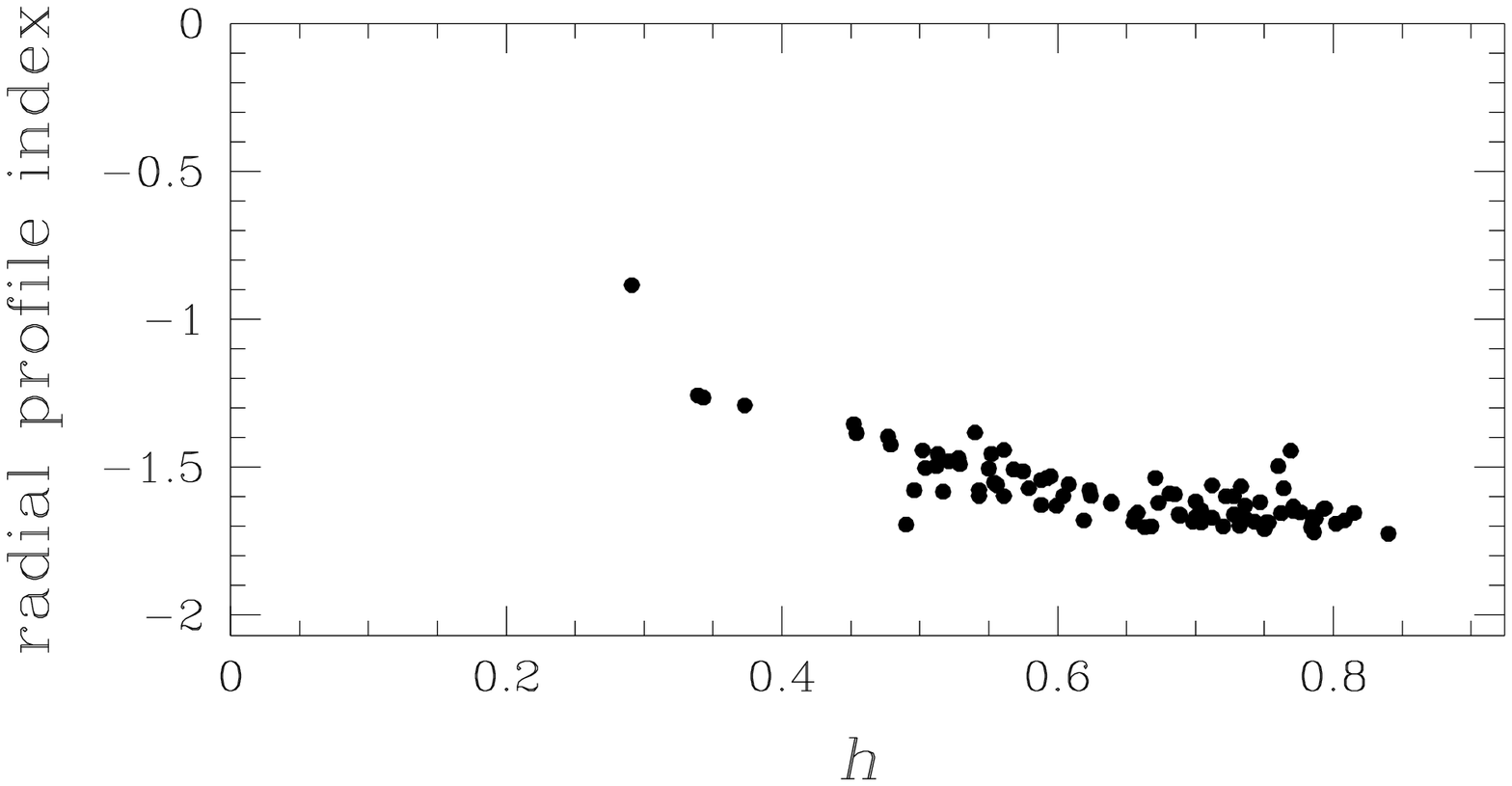}
\includegraphics[width=8.5cm]{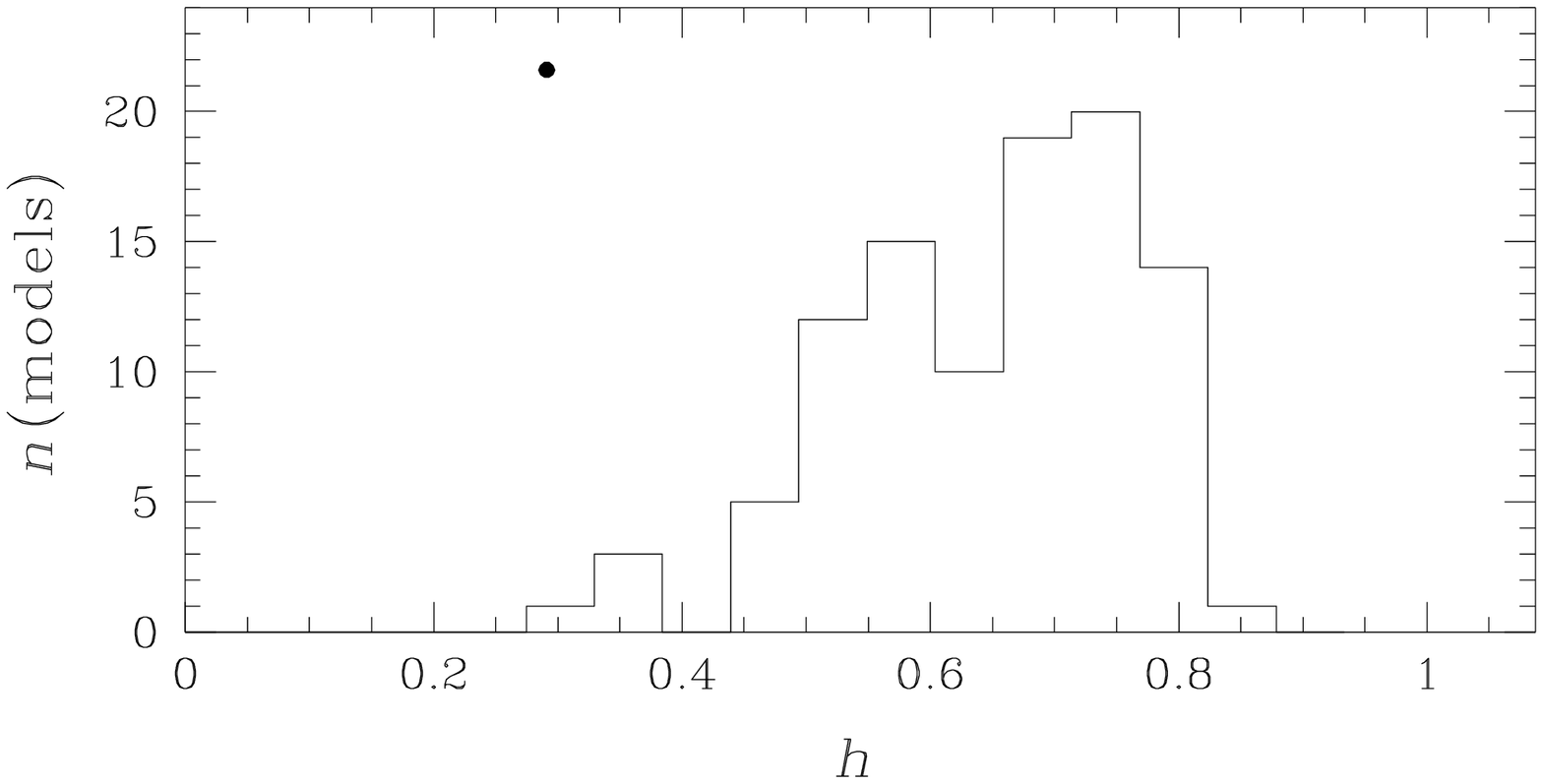}
\caption{{\it Top:} radial index profile vs. H$_0$ for
one hundred different non-parametric models, showing that all the
 galaxy models have a steep mass profile ($<$ -0.5)
 {\it Bottom:}
probability density for H$_0$ for the same models. Units on the $X$
axis of both figures is in $h_{100}$. The median value of the
distribution is 65 $\ho$. The black dot marks the location of an isothermal
mass profile (radial profile index = -1).}
\label{hr_saha}
\end{figure}

\subsection{Pixelated models}

The   pixelated    models   of   Saha    \&   Williams   (\cite{Saha})
and Williams \& Saha (\cite{will_saha}) were  also applied to  \object{HE~2149$-$2745}.  While
the  mass distribution  resulting from  these  models is  not as  well
related to the physical parameters  of the lens, they allow us to explore
a wide  range of  lens shapes  and to estimate  the robustness  of the
time-delay    conversion.   Fig.~\ref{mass_saha}   shows    the   mass
distribution found by  the models. During the fit,  the ellipticity of
the mass  distribution was kept free. The  program reconstructs almost
round    mass   distributions,   as    does   the    analytical   model
(Fig.~\ref{cut_ell}). We  also examine the  impact of a change  in the
compactness   of  the   mass  profile   on  the   derived   value  for
H$_0$. Fig.~\ref{hr_saha} shows the result of a Monte-Carlo simulation
where 100 models were run  with different indices for the lens shape.
The  density  index in  Fig.~\ref{hr_saha}  represents the  logarithmic
projected density gradient in the vicinity of the images (cf. Williams
\& Saha, \cite{will_saha}).  The upper  panel of the figure shows that
most models predict relatively steep/concentrated mass profiles.

In  the  lower panel  of  Fig.~\ref{hr_saha}  we  display the  H$_0$
probability distribution. The mean of the distribution peaks at H$_0 =
65 \pm 15\, \ho$ at the 2-$\sigma$  level and H$_0 = 65 \pm 8\, \ho$ at
the 1-$\sigma$ level.

\section{Summary - conclusions}
\label{sect:discussion}

We have presented the first result of a long-term photometric
monitoring campaign undertaken at ESO between 1998 and 2000 at the
1.54-m Danish telescope. Our $V$ and $i$ light curves allow us to
measure a time-delay of $\Delta t = 103 \pm12$ days between the two 
quasar images of \object{HE~2149$-$2745}.

From VLT  spectroscopy, we  have derived a  tentative estimate  of the
lens redshift  to be $z=0.489$.  Applying both  analytic and numerical
lens models to  the case of \object{HE~2149$-$2745}, we  derive H$_0 =
66 \pm 6\,  \ho$ with an additional systematic error  of $\pm\, 3\, \ho$
in  the  case  of  the  analytic  models, due  the  limited  range  of
the ellipticity. The derived  mass models are relatively compact,
although not as  compact as the light profile of  the galaxy. An extra
source of  systematics might be  introduced by the uncertainty  on the
lens redshift estimate.  This is  however not critical for the
determination of H$_0$ as the error is dominated by the uncertainty in
the  time-delay measurement.  As \object{HE~2149$-$2745}  shows smooth
light  curves, it  is likely  that the  situation can  be  improved by
continued monitoring.  With an  improved time-delay it would also  be highly 
desirable to re-determine the lens
redshift  more precisely.

Our monitoring program of  \object{HE~2149$-$2745} is the first to be carried out in two bands on
such a regular basis and over such a long time scale. Given the error
bars, we do not see any significant colour variation over the 900 days of
observation. Moreover, our spectra of the two quasar images show that
the flux ratios in the broad emission lines behave differently for the
continuum flux ratio, and that the flux ratio measured in the BAL
structure of the source follows the behavior as the continuum region.
Such behavior can be explained both by microlensing or by differential 
extinction
by the lensing galaxy, or both.  So far, the data do not allow us to
distinguish between the two possible explanations.  In order to
confirm microlensing, one would need for example to know the
time-scale of the putative event, i.e., to measure the absolute
magnification of the event and its duration.  This would not only
allow us to confirm microlensing, but also to use it to map the radial
structure of the central AGN in the source. 
Although the  Einstein radius
crossing time is long for \object{HE~2149$-$2745}, of the order of 10 years,
much shorter microlensing events, such as a caustic crossings, can occur.
Conducting a long-term spectrophotometric monitoring could  
therefore allow us to probe the AGN size \object{HE~2149$-$2745}.

\begin{acknowledgements}
We thank IJAF and ESO for  granting us observing time for this project
on a flexible basis.  We are  very grateful to the 2.2-m team for their
support and enthusiasm. Most  of the monitoring observations presented
here  have been  carried out  during short  periods of  time allocated
during the  scheduled runs of  regular observers. We  appreciated very
much the enthusiasm  of all the observers who  accepted to perform the
observations for us.  Some of them, present at  the telescope for long
periods, contributed a  lot to our program. It is  a pleasure to thank
in  particular  B.   Milvang-Jensen,  L.  Fogh-Olsen,
S.  Frandsen, L.  Hansen, H.  Kjeldsen, J.  Knude
and G.   Israel. 
We  thank  Lutz
Wisotzki for letting us use  his EFOSC spectrum and Prasenjit Saha for
help  with the  non-parametric  models.  IB was  supported  by   
P\^ole  d'Attraction
Interuniversitaire,  P4/05 \protect{(SSTC, Belgium)}.  FC acknowledges
financial support through  Chilean grant FONDECYT/3990024.  Additional
support from  the European Southern  Observatory, through ECOS/CONICYT
grant  C00U05, CNRS/CONICYT grant 8730 and Marie Curie grant 
MCFI-2001-00242 are  also gratefully
acknowledged.  JH is supported  by  the Danish Natural  Science Research  
Council  (SNF). JPK thanks CNRS for support.
\end{acknowledgements}

\end{document}